\documentclass[runningheads]{llncs}
\usepackage{times} 
\usepackage{graphicx}
\usepackage{todonotes}
\usepackage{tikz}
\usepackage{listings}
\usepackage{algorithm2e}
\usepackage{enumitem}
\usepackage{wrapfig}
\usepackage{url}
\usepackage{ctable} 
\usepackage{multirow}
\usepackage{booktabs}
\usepackage{float}
\usepackage{wrapfig} 
\usepackage{caption}
\usepackage{comment}
\usepackage{amsmath}
\usepackage{amssymb}
\usepackage{subcaption}
\definecolor{darkgreen}{rgb}{0.0, 0.3, 0.0}
\definecolor{codegreen}{rgb}{0,0.4,0}
\definecolor{codegray}{rgb}{0.5,0.5,0.5}
\definecolor{codepurple}{rgb}{0.58,0,0.82}
\definecolor{darkmagenta}{rgb}{0.55,0,0.55}
\definecolor{backcolour}{rgb}{1.0,1.0,1.0}

\newcommand{\uclid}{UCLID5\xspace}
\newcommand{\boogie}{Boogie\xspace}

\newcommand{\smtlib}{SMT-LIB\xspace}
\newcommand{\llamalib}{Synth-Lib\xspace}
\newcommand{\sygusif}{SyGuS-IF\xspace}
\newcommand{\tap}{{\textsc{TAP}}\xspace}

\newcommand{\codelike}[1]{\texttt{#1}}

\lstdefinelanguage{uclid}{
  sensitive = true,
  keywords={module, forall, exists, Lambda, if, else, assert, assume, havoc,
            for, range, skip, case, esac, init, next, control, function, procedure, oracle,
            returns, call, define, type, var, input, output, const, property,
            invariant, synthesis, grammar, requires, ensures, modifies, instance, axiom, 
            enum, record, integer, boolean, float, true, false, finite_forall, finite_exists, group},
  numbers=left,
  numberstyle=\footnotesize,
  stepnumber=1,
  numbersep=8pt,
  showstringspaces=false,
  breaklines=true,
  frame=top,
  comment=[l]{//},
  morecomment=[s]{/*}{*/},
}

\lstdefinelanguage{smt}{
  sensitive = false,
  keywords={declare, fun, synth, check, assert, sat, blocking, define, constraint},
  numbers=left,
  numberstyle=\footnotesize,
  stepnumber=1,
  numbersep=8pt,
  showstringspaces=false,
  breaklines=true,
  frame=top,
  comment=[l]{;},
}

\lstdefinestyle{uclidstyle}{
  backgroundcolor=\color{backcolour},
  commentstyle=\color{codegreen},
  keywordstyle=\color{magenta},
  numberstyle=\tiny\color{codegray},
  stringstyle=\color{codepurple},
  basicstyle=\footnotesize,
  breakatwhitespace=false,
  basicstyle=\bf\scriptsize\ttfamily,
  breaklines=true,
  captionpos=b,
  keepspaces=true,
  numbers=left,
  numbersep=5pt,
  showspaces=false,
  showstringspaces=false,
  showtabs=false,
  tabsize=2,
  frame=shadowbox
}

\newcommand{\mat}[1]{{#1}}
\newif\ifcontrol
\controltrue

\newcommand{\hide}[1]{} 

\usepackage{graphicx}                   
\usepackage{hyperref}                   
\usepackage[firstpage]{draftwatermark}  

\SetWatermarkAngle{0}

\SetWatermarkText{
\raisebox{10.0cm}{%
\hspace{0.1cm}%
  \href{https://doi.org/10.1109/5.771073}{\includegraphics{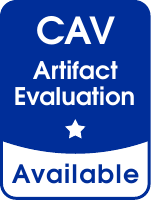}}%
  \hspace{9cm}%
  \includegraphics{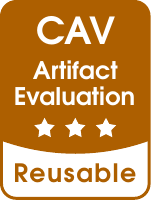}%
}}

\begin{document}
\title{\uclid: Multi-Modal Formal Modeling, Verification, and Synthesis}
\titlerunning{\uclid}
\authorrunning{Polgreen et al}
%
\author{Elizabeth Polgreen \inst{1,2}\orcidID{0000-0001-9032-7661}
 \and 
Kevin Cheang\inst{1}\orcidID{0000-0002-5717-0575}
\and
Pranav Gaddamadugu\inst{1}
\and
Adwait Godbole \inst{1}\orcidID{0000-0001-7704-304X}
\and
Kevin Laeufer\inst{1}\orcidID{0000-0003-0942-7070}
\and
Shaokai Lin\inst{1}\orcidID{0000-0001-6885-5572}
\and
Yatin A.~Manerkar\inst{3}
\and
Federico Mora\inst{1}\orcidID{0000-0002-0725-9213}
\and Sanjit A.~Seshia\inst{1}\orcidID{0000-0001-6190-8707}}
%

%
\institute{UC Berkeley \and University of Edinburgh \and University of Michigan
}
\maketitle              
\begin{abstract}
\uclid is a tool for the multi-modal formal modeling, verification, and synthesis of systems.
It enables one to tackle verification problems for heterogeneous systems such as combinations of hardware and software,
or those that have multiple, varied specifications, 
or systems that require hybrid modes of modeling.
A novel aspect of \uclid is an emphasis on the use of syntax-guided and inductive synthesis to automate steps in modeling and verification. 
This tool paper presents new developments in the \uclid tool including 
new language features, 
integration with new techniques for syntax-guided synthesis and satisfiability solving,
support for hyperproperties and combinations of axiomatic and operational modeling,
demonstrations on new problem classes, 
and a robust implementation.

%
\end{abstract}
\section{Overview} 

Tools for formal modeling and verification are typically specialized for particular domains and for particular methods.
For instance, software verification tools like Boogie~\cite{boogie} focuses on modeling sequential software and Floyd-Hoare style reasoning, while 
hardware verifiers like ABC~\cite{abc} are specialized for sequential circuits and SAT-based equivalence and model checking.
Specialization makes sense when the problems fit well within a homogeneous problem domain with specific verification needs.
However, there is an emerging class of problems, such as in security and cyber-physical systems (CPS), where the systems under verification are heterogeneous, or the types of specifications to be verified are varied, or there is not a single type of model that is effective for verification.
An example of such a problem is the verification of trusted computing platforms~\cite{subramanyan-ccs17} that involve hardware and software components working in tandem, and where the properties to be checked include invariants, refinement checks, and hyperproperties.
There is a need for automated formal methods and tools to handle this class of problems.

\uclid is a system for {\em multi-modal} formal modeling, verification, and synthesis that addresses the above need.
\uclid is multi-modal in three important ways.
First, it permits different modes of modeling, using axiomatic and
operational semantics, or as combinations of concurrent transition systems and procedural code.
This enables modeling systems with multiple characteristics.
Second, it offers a varied suite of specification modes,
including first-order formulas in a combination of logical theories,
temporal logic, inline assertions, pre- and post-conditions, system invariants,
and hyperproperties.
Third, it supports the first two capabilities with a varied suite of verification techniques, including Floyd-Hoare style proofs, k-induction and bounded model checking (BMC), verifying hyperproperties,
or using syntax-guided and inductive synthesis to provide more automation in tedious steps of verification, or to automate the modeling process (as proposed in~\cite{seshia-pieee15}).


The \uclid framework was first proposed in 2018~\cite{memocode}, itself a major evolution of the much older UCLID system~\cite{bryant-cav02}, one of the first satisfiability modulo theories (SMT) based modeling and verification tools.  
Since that publication~\cite{memocode}, which laid out the vision for the tool and described a preliminary implementation, the utility of the tool has been demonstrated on several problem classes (e.g.,~\cite{cheang-csf19,magyar-iccad19,Cheang2020VerifyingRP}),
such as for verifying security across the hardware-software interface. 
The syntax has been extended and state-of-the-art methods for syntax-guided synthesis (SyGuS) have also been integrated into the tool~\cite{mora-arXiv20}, including new capabilities for satisfiability and synthesis modulo oracles~\cite{symo}.
This tool paper presents an overview of the latest version of \uclid, highlighting novel multi-modal aspects of the tool, as well as the new features supported since 2018~\cite{memocode}. The paper is structured as follows: in Section ~\ref{sec:overview} we give an overview of the \uclid tool; in Section~\ref{sec:features} we detail different multi-modal aspects of the tool, as well as high-lighting new features; and in Section~\ref{sec:casestudies} we present a case study using \uclid to verify a Trusted Abstract Platform. We cover related work in Section~\ref{sec:related}. 
The new features we highlight are:
\begin{enumerate}
    \item Fully integrated support for synthesis across all verification modes
    \item Support for modeling with external oracles, via satisfiability and synthesis modulo oracles~\cite{symo}
    \item New language features to support combining axiomatic and operational modeling
    \item Direct support for hyperproperties
    \item Front-end translations from Chisel/FIRRTL to \uclid, and from RISC-V binaries to \uclid, referenced in Section~\ref{sec:software}.
    \item New case studies: covering models for distributed CPS in Lingua Franca~\cite{linguafranca}, and encodings of $\mu$hb specifications and verification of a Trusted Abstract Platform described in Sections~\ref{sec:operAx} and~\ref{sec:tap} and in the corresponding artifact~\cite{artifact}.
\end{enumerate}

\section{Overview of \uclid}
\label{sec:overview}
In verification mode, \uclid reduces the question of whether a model satisfies a given specification to a set of constraints that can be solved by an off-the-shelf SMT solver. In synthesis mode, \uclid reduces the problem of finding an interpretation for an uninterpreted function such that the specification is satisfied into a SyGuS problem that can be solved by an off-the-shelf SyGuS solver. In order to do so, \uclid performs the following main tasks, as shown in Figure~\ref{fig:flow}:

\paragraph{Front end:} \uclid takes models written in the \uclid language as input. The command-line front-end allows user configuration, including specifying the external SMT-solver/SyGuS-solver to be used, as well as enabling certain utilities such as automatically converting uninterpreted functions to arrays. 
The parser builds an abstract syntax tree from the model.

\paragraph{AST passes:} \uclid performs a number of transformations and checks on the abstract syntax tree, including type-checking and inlining of procedures. This intermediate representation supports limited control flow such as if-statements and switch-cases, but loops are not permitted in procedural code and are removed via unrolling (bounded for-loops) or replacement with user-provided invariants (while loops). However, unbounded control flow can be handled by representation as transition systems (where each module consists of a transition system with an initial and a next block, each represented as a separate AST).

\paragraph{Symbolic Simulator:} The symbolic simulator performs a simulation of the transition system in the model, according to the verification command provided, and produces a set of assertions. For instance, if bounded model checking is used, \uclid will symbolically execute the main module a bounded number of times. 
 \uclid encodes the violation of each independent verification condition as a separate assertion tree.

\paragraph{\llamalib interface:} \uclid supports both synthesis and verification. The \llamalib interface constructs either a verification or a synthesis problem from the assertions generated by the symbolic simulator. The verification problems are passed to the \smtlib interface, which converts each assertion in \uclid's intermediate representation to an assertion in \smtlib. Similarly, the synthesis problems are passed to the \sygusif interface, which converts each assertion to an assertion in \sygusif. The verification and synthesis problems are then passed to the appropriate provided external solver and the result is reported back to the user. 

\begin{figure}
    \centering
    \begin{tikzpicture}[>=latex,x=3cm,y=2cm]
\node[rectangle,draw,minimum height=0.9cm, minimum width=1.7cm ,align=center, rounded corners] at (0,0) (f) {Front-end\\ parser};
\node[rectangle,draw,minimum height=0.7cm,minimum width=1.8cm,align=center, rounded corners] at (0.9,0) (ast) {AST \\transformation \\passes};
\node[rectangle,draw,minimum height=0.7cm,minimum width=1.8cm,align=center, rounded corners] at (1.8,0) (ss) {Symbolic\\ Simulator};
\node[rectangle,draw,minimum height=0.7cm,minimum width=1.8cm,align=center, rounded corners] at (2.7,0) (synthlib) {\llamalib\\ interface};
\node[rectangle,draw,minimum height=0.7cm,minimum width=1.8cm,align=center, rounded corners] at (2.25,-0.9) (smt) {\smtlib\\ interface};
\node[rectangle,draw,minimum height=0.7cm,minimum width=1.8cm,align=center, rounded corners] at (3.15,-0.9) (sygus) {\sygusif\\ interface};
\node[] at (2.25, -1.7) (smtsolver){SMT solver};
\node[] at (3.15, -1.7) (sygussolver){SyGuS solver};
\node[align=center, minimum width=1cm] at (3.4, 0) (cex){Result + \\c-example};

\path[->] (f) edge[above] node {AST} (ast);
\path[->] (ast) edge[above] node {AST} (ss);
\path[->] (ss) edge[above, align=center] node {assert\\tree} (synthlib);
\path[->] (synthlib) edge[align=center, right] node {synth\\IR} (sygus);
\path[->] (synthlib) edge[align=center, left] node {synth\\IR} (smt);
\path[->] (smt) [bend left, right] edge node {query} (smtsolver);
\path[->] (smtsolver) [bend left, left] edge node {model} (smt);
\path[->] (sygus)[bend left, right]  edge node {query} (sygussolver);
\path[->] (sygussolver) [bend left, left] edge node {model} (sygus);
\path[->] (synthlib) edge node {} (cex);



\end{tikzpicture}



    \caption{Architecture of \uclid}
    \label{fig:flow}
\end{figure}

\subsubsection{Basic \uclid Models}
A simple \uclid model that computes the Fibonacci sequence is shown in Figure~\ref{ex:fib}. \uclid models are contained within modules which comprise of $3$ parts: a system model represented using combinations of sequential, concurrent, operational and axiomatic modeling, as described in Sections~\ref{sec:operAx}; a system specification described in Section~\ref{sec:verification_modes}; and a proof script that specifies the verification tasks \uclid should perform to prove that the system satisfies its specification, using a variety of supported verification and synthesis techniques described in Section~\ref{sec:verification_modes}.

\section{Multi-modal Language Features} 
\label{sec:features}
\subsection{Multi-modal verification and synthesis}

\vspace*{-2mm}
\subsubsection{Specification}
\label{sec:verification_modes}
\uclid supports a variety of different types of specifications. The standard properties supported include inline assertions and assumptions in sequential code, pre-conditions and post-conditions for procedures, and global axioms and invariants (both as propositional predicates, and temporal invariants in Linear Temporal Logic (LTL)). 

The latest version of \uclid further provides direct support for hyperinvariants and hyperaxioms (for $k$-safety).
This new support for direct hyperproperties comprises of two new language constructs: \lstinline{hyperaxiom} and \lstinline{hyperinvariant}. The former places an assumption on the behavior of the module, if $n$ instances of the module were instantiated, and the latter is an invariant over $n$ instances of the module, which is verified via the usual verification methods. A variable $x$ from the $n^{th}$ instance of the module is reasoned about in the predicate using $x.n$, and the number of modules instantiated is determined by the maximum $n$ in both the invariant and the axiom. For example, \codelike{hyperinvariant[2] det\_xy: y.1 == y.2} asserts that a $2$-safety hyperproperty holds. 

\vspace*{-2mm}
\subsubsection{Verification}

To verify these specifications, we implement multiple classic  techniques. As a result, once a model is written in \uclid, the user can deploy a combination of verification techniques, depending on the properties targeted. \uclid supports a range of verification techniques including: Bounded Model Checking (for LTL, hyperinvariants and assertion-based properties); induction and k-induction for assertion-based invariants and hyperinvariants; and verification of pre-and post-conditions on procedures and hyperinvariants.

As an exemplar of the utility of multi-modal verification, consider the hyper-property based models verified by Sahai et al.~\cite{hyperproperties}. These models use both procedure verification and induction to verify k-trace properties.


\vspace*{-2mm}
\subsubsection{Synthesis}

\label{sec:synthesis}
The latest version of \uclid integrates program synthesis fully across all the verification modes previously described. Specifically, users are able to declare and use \emph{synthesis functions} anywhere in their models, and UCLID5 will seek to  automatically synthesize function bodies for these functions such that the user-selected verification task will pass. 
In this section, we give an illustrative example of synthesis in \uclid, we provide the necessary background on program synthesis, and then we formulate the existing verification techniques inside of \uclid for synthesis.

Consider the \uclid model in Fig.~\ref{ex:fib}. The user wants to prove by induction that the invariant \codelike{a\_le\_b} at line 13 always holds. Unfortunately, the proof fails because the invariant is not inductive. Without synthesis, the user would need to manually strengthen the invariant until it became inductive. However, the user can ask \uclid to automatically do this for them. Fig.~\ref{ex:fib} demonstrates this on lines 16, 17 and 18. Specifically, the user specifies a function to synthesize called \codelike{h} at lines 16 and 17, and then uses \codelike{h} at line 18 to strengthen the existing set of invariants. Given this input, \uclid, using e.g. \textsc{cvc5}~\cite{cvc4} as a syntax-guided synthesis engine, will automatically generate the function \codelike{h(x, y) = x >= 0}, which completes the inductive proof. 

In this example, the function to synthesize represents an inductive invariant. However, functions to synthesize are treated exactly like any interpreted function in \uclid: the user could have called \codelike{h} anywhere in the code. Furthermore, this example uses induction and a global invariant, however, the user could also have used a linear temporal logic (LTL) specification and bounded model checking (BMC). In this sense, our integration is fully flexible and generic.
Furthermore, the integration scheme allows us to enable synthesis for any verification procedure in \uclid, by simply letting users declare and use functions to synthesize and relying on existing \sygusif solvers to carry out the automated reasoning.

\begin{figure}[t]
\begin{lstlisting}[language=uclid,style=uclidstyle]
module main {

  // Part 1: System Description.
  var a, b : integer;
  init {
    a, b = 0, 1;
  }
  next {
    a', b' = b, a + b;
  }

  // Part 2: System Specification.
  invariant a_le_b: a <= b; 
  
  // Part 3: (NEW) Synthesis Integration
  synthesis function 
    h(x : integer, y : integer): boolean;
  invariant hole: h(a, b);

  // Part 4: Proof Script.
  control {
    induction;
    check;
    print_results;
  }
}
\end{lstlisting}
  \caption{\uclid Fibonacci model. Part 3 shows the new synthesis syntax, and how to find an auxiliary invariant.}
  \label{ex:fib}
\end{figure}

\subsection{Multi-modal modeling}

\subsubsection{Combining Concurrent and Sequential Modeling}
\label{sec:seq-concur}
A unique feature of the \uclid modeling language is the ability to easily combine sequential and concurrent modeling. This allows a user to easily express models representing sequential programs, including standard control flow, procedure calls, sequential updates, etc, in a sequential model, and to combine these components within a system designed for concurrent modeling based on transition systems. The sequential program modeling is inspired by systems such as Boogie~\cite{boogie} and allows the user to port Boogie models to \uclid. The concurrent modeling is done by defining transition systems with a set of initial states and a transition relation. Within \uclid, each module is a transition system. A main module can be defined that triggers when each child module is stepped. 
For an example of this combination of sequential and concurrent modeling, we refer the reader to the CPU example presented in the original \uclid paper~\cite{memocode}, which uses concurrent modules to instantiate multiple CPU modules, modeled as transition systems, with sequential code to model the code that executes instructions, and to the case study in Section~\ref{sec:tap}.

\subsubsection{Reasoning with External Oracles}

\label{sec:smto}
New in the latest version,
\uclid supports the modeling with \emph{oracle function symbols}~\cite{symo} in both verification and synthesis.
 Namely, a user can include ``oracle functions'' in any \uclid model, where an oracle function is a function without a provided implementation, but which is associated to a user-provided external binary that can be queried by the solver. We note that oracle functions (and functions in general) can only be first-order within the UCLID5 modeling language, i.e., functions cannot receive functions as arguments. 
 
 This support is useful in cases where some components of the system are difficult or impossible to model, but could be compiled into a binary that the solver can query; or where the model of the system would be challenging for an SMT solver to reason about (for instance, highly non-linear arithmetic), and it may be better to outsource that reasoning to an external binary.
 
 \uclid supports oracle function symbols in verification by interfacing with a solver that supports Satisfiability Modulo Theories and Oracles (SMTO)~\cite{symo}, and in synthesis by interfacing with a solver that supports Synthesis Modulo Oracles (SyMO)~\cite{symo}.
 
  Oracle function symbols are declared like functions, with the keyword \texttt{oracle}, and an annotation pointing to the binary implementation. For instance \texttt{oracle function [isprime] Prime (x: integer) : boolean} would indicate to the solver that the binary \texttt{isprime} takes an integer as input and returns a boolean. This is translated into the corresponding syntax in SMTO or SyMO, as detailed in~\cite{sygus-if}.

 %
%
%
%
%
%
An exemplar of such reasoning in a synthesis file is available in the artifact~\cite{artifact}, where we use \uclid to synthesize a safe and stabilizing controller for a Linear Time Invariant system, similar to Abate et al.~\cite{control-cegis}.

\subsubsection{Combining Operational and Axiomatic Modeling}
\label{sec:operAx}
\uclid can model a system being verified using an operational (transition system-based) approach, as Figure~\ref{ex:fib} shows. However, \uclid also supports modeling a system in an \emph{axiomatic} manner, whereby the system is specified as a set of properties over traces. Any execution satisfying the properties is allowed by the system, and any execution violating the properties is disallowed. Axiomatic modeling can provide order-of-magnitude performance improvements over operational models in certain cases~\cite{alglave:herding}, and is often well suited to systems with large amounts of non-determinism. We provide an example of fully axiomatic modeling in the artifact~\cite{artifact}. 

However, uniquely, \uclid allows users to specify multi-modal systems using a combination of operational and axiomatic modeling. In such models, some constraints on the execution are enforced by the initial state and transition relation (operational modeling), while others are enforced through axiomatic invariants (axiomatic modeling). This allows the user to choose the mode of modeling most appropriate to each constraint.
For example, the ILA-MCM work~\cite{zhang:ilamcm} combined operational ILA (Instruction Level Abstraction) models to describe the functional behavior of processing elements with memory consistency model (MCM) orderings that are more naturally specified axiomatically~\cite{alglave:herding}. (MCM orderings constrain shared-memory communication and synchronization between multiple processing elements.)
The combined model, used for System-on-Chip verification, worked by sharing variables (called ``facets'') between both the models.
\uclid makes it much easier to perform such a combination.

\begin{figure}[!ht]
\begin{lstlisting}[language=uclid,style=uclidstyle]
module main {
    <type declarations>
    var i1, i2, i3, i4 : microop_t;
    <set i1-i4 to be the instructions of a test, like mp>
    group testInstrs : microop_t = {i1, i2, i3, i4};
    
    //Vars to decide which instrs to step and when.
    var next1, next2, next3, next4 : boolean;
    var curTime : integer;

    init {
        i1.Fetch.nExists = false; i1.Execute.nExists = false;
        <...>
    }
    //Axiom enforcing that instructions are fetched in order.
    axiom fifoFetch :
        finite_forall (i : microop_t) in testInstrs ::
        finite_forall (j : microop_t) in testInstrs ::
        (ProgramOrder(i, j) && NodeExists(j.Fetch)) ==> 
            EdgeExists(i.Fetch, j.Fetch);

    procedure stepInst(index : integer)
      returns (instr_next : microop_t)
    {
        //Steps instr@index, unless it has completed.
        case
            (index == 1) : {
                instr_next = i1;
                if(!instr_next.Fetch.nExists) {
                    instr_next.Fetch.nExists = true; 
                    instr_next.Fetch.nTime = curTime;
                } else {
                <...>
        esac
    }
    next {
        //Increment the current timestamp and
        //nondeterministically step instructions.
        curTime' = curTime + 1;
        havoc next1, next2, next3, next4;
        
        if (next1) { call (i1') = stepInst(1); }
        if (next2) { call (i2') = stepInst(2); }
        if (next3) { call (i3') = stepInst(3); }
        if (next4) { call (i4') = stepInst(4); }
    }
}
\end{lstlisting}
  \vspace{-10pt}
  \caption{\uclid model that incorporates both operational modeling (through the \texttt{init} and \texttt{next} blocks) and axiomatic modeling (through the \texttt{axiom} keyword).}
  \label{fig:operAx}
\end{figure}

Fig~\ref{fig:operAx} depicts parts of a \uclid{} model of microarchitectural execution that uses both operational and axiomatic modeling (similar to that from the ILA-MCM work), 
based on the $\mu$spec specifications of COATCheck~\cite{lustig:coatcheck}.
In this model, the steps of instruction execution are driven by the \texttt{init} and \texttt{next} blocks, i.e., the operational component of the model. Multiple instructions can step at any time (\texttt{curTime} denotes the current time in the execution), but they can only take one step per timestep.
Meanwhile, axioms such as the \texttt{fifoFetch} axiom enforce ordering \emph{between} the execution of multiple instructions.
The \texttt{fifoFetch} axiom specifically enforces that instructions in program order on the same core must be fetched in program order.
(Enforcing this order is tricky using operational modeling alone). 
The transition rules and axioms operate over the same data structures, ensuring that executions of the final model abide by both sets of constraints.

$\mu$spec models routinely function by grounding quantifiers over a finite set of instructions. Thus, to fully support $\mu$spec axiomatic modeling, we introduce two new language features ---namely, {\em groups} and {\em finite quantifiers}. A group is a set of objects of a single type. A group can have any number of elements, but it must be finite, and the group is immutable once created. For instance, the group \texttt{testInstrs} in Figure~\ref{fig:operAx} consists of four instructions. 
Finite quantifiers, meanwhile, are used to quantify over group elements.


This example showcases \uclid's highly flexible multi-modal modeling capability. Models can be purely operational, purely axiomatic, or a combination of the two. Note that axiomatic modeling relies on the new language features \texttt{finite\_forall} and \texttt{groups}. 
For a further example of axiomatic and operational multi-modal modeling, we refer the reader to the case study checking reachability properties in reactive embedded systems described in the artifact~\cite{artifact}.

\section{Case Study: TAP model}
\label{sec:casestudies}
\begin{figure}[htbp]
\begin{lstlisting}[language=uclid,style=uclidstyle]
module tap {
  // State variable declarations
  var tap_enclave_metadata_valid: tap_enclave_metadata_valid_t;
  var tap_enclave_metadata_addr_map: tap_enclave_metadata_addr_map_t;
  <...>
  
  // Enclave operations
  procedure launch(eid: tap_enclave_id_t, <...>) { <...> }
  <...>
  
  init { <...> } // initialize TAP
  
  next { // step the system
    case
      (tap_current_mode == mode_untrusted) : {
        call (<...>) = AdversarialStep(<...>);
      }
      (tap_current_mode == mode_enclave) : {
        call (<...>) = EnclaveStep(<...>);
      }
    esac
  }
}
  
module integrity_proof {
  // Create two instances of the TAP model 
  instance tap_1: tap(<...>);
  instance tap_2: tap(<...>);
  
  // Example invariant: Memory that is mapped are equal between the two traces
  invariant equal_mem: (forall (pa : wap_addr_t) ::
    e_excl_map[pa] ==> (tap_1.mem[pa] == tap_2.mem[pa]));
  <...>
  
  init { <...> } // initialize proof
  
  next { // step the system
    next(tap_1); next(tap_2);
  }
  
  control {
    v = induction;
    check;
  }
}
\end{lstlisting}
  \caption{\uclid{} transition system-styled model of \tap and the integrity proof.}
  \label{fig:tap}
\end{figure}

\label{sec:tap}
The final case study we wish to describe verifies a model of a trusted execution environment. 
Trusted execution environments \cite{keystone,mitsanctum,intelsgx,tdx} often provide a software interface for users to execute enclaves, using hardware primitives to enforce memory isolation. In contrast to software which requires reasoning about sequential code, hardware modeling uses a paradigm that permits concurrent updates to a system. Moreover, verifying hyperproperties such as integrity requires reasoning about multiple instances of a system which most existing tools are not well suited for. In this section, we present the \uclid{} port~\footnote{\url{https://github.com/uclid-org/trusted-abstract-platform/}} of the Trusted Abstract Platform (\tap) which was originally~\footnote{\url{https://github.com/0tcb/TAP}} written in \boogie{} and introduced by Subramanyan et. al. \cite{subramanyan-ccs17} to model an abstract idealized trusted enclave platform. We demonstrate how \uclid{}'s multi-model support alleviates the difficulties in modeling the \tap model in existing tools.

\subsubsection{Modeling the \tap and Proving Integrity}
The \uclid{} model of \tap in Figure~\ref{fig:tap} demonstrates some of \uclid{}'s key features: the enclave operations of the \tap model 
(e.g. \codelike{launch}) 
are implemented as procedures, and a transition relation of the \tap is defined using a next block that either executes an untrusted adversary operation or the trusted enclave, which in turn executes one of the enclave operations atomically. Proving the integrity hyperproperty on the \tap thus only requires two instantiations of the \tap model, specifying the integrity invariants, and defining a next block which steps each of the \tap instances as shown in the \codelike{integrity\_proof} module. The integrity proof in \uclid{} uses inductive model checking.

\subsubsection{Results and statistics of the \tap modules}
\begin{wrapfigure}{R}{0.55\linewidth}
    \vspace{-20pt}
    \begin{tabular}{ p{0.2\textwidth} p{0.05\textwidth} p{0.05\textwidth} p{0.05\textwidth} p{0.05\textwidth} c }
    \specialrule{.1em}{.1em}{.1em} 
     \multirow{2}{*}{\textbf{Model/Proof}} & \multicolumn{4}{c}{\centering\textbf{Size}} & \multirow{2}{*}{\vtop{\hbox{\strut \textbf{Verif.}}\hbox{\strut \textbf{Time (s)}}}} \\
     & \#pr & \#fn & \#an & \#ln\\
     \hline
     \multicolumn{6}{c}{\textbf{Boogie}}\\
     \hline
     \texttt{TAP} & 22 & 25 & 254 & 1840 & 51 \\
     \texttt{Integrity} & 14 & 11 & 71 & 835 & 346\\
     \hline
     \multicolumn{6}{c}{\textbf{\uclid{}}}\\
     \hline
     \texttt{TAP} & 53 & 25 & 87 & 2765 & 49 \\
     \texttt{Integrity} & 2 & 0 & 54 & 293 & 30\\
    \specialrule{.1em}{.1em}{.1em}
    \end{tabular}
    \captionof{table}{\boogie{} vs \uclid{} Model Results\label{table:boogievsuclid5}}
    \vspace{-20pt}
\end{wrapfigure}

Table~\ref{table:boogievsuclid5} shows the approximate size of the \tap model in both \boogie{} and \uclid{}. \#pr, \#fn, \#an, and \#ln refer to the number of procedures, functions, annotations, and lines of code respectively. Annotations are the number of loop invariants, assertions, assumptions, pre- and post-conditions that were manually specified. The verification time includes compilation and solving.



While the \#ln for the \tap model in \uclid{} is higher than that of the model in \boogie{} due to stylistic differences, the crucial difference is in the integrity proof. The original model in \boogie{} implements the \tap model and integrity proof as procedures, where the transition of the \tap model is implemented as a while loop. However, this lack of support for modeling transition systems introduces duplicate state variables in a hyperproperty such as integrity, requires context switching and additional procedures for the new variables, which makes the model difficult to maintain and self composition unwieldy. In \uclid{}, the proof is no longer implemented as a procedure, but rather, we create instances of the \tap model. We also note that the number of annotations is less in \uclid{} compared to \boogie{} for the \tap model and proof. Additionally, 
this model lends itself for more direct verification of hyperproperties.

The verification results are run on a machine with 2.6GHz 6-Core Intel Core i7 and 16GB of RAM running OSX. As shown on the right of Table~\ref{table:boogievsuclid5}, the verification runtimes between the \boogie{} and \uclid{} models and proofs are comparable.

\WFclear



\section{Related Work} 
\label{sec:related}




There are a multitude of verification and synthesis tools related to \uclid{}. In this brief review, we highlight prominent examples and contrast them with \uclid{} along the key language features described in Section~\ref{sec:features}. 

\uclid{} allows users to combine sequential and concurrent modeling (see Section~\ref{sec:seq-concur}). Most existing tools primarily support either sequential, e.g. \cite{dafny,boogie,rosette}, or concurrent computation modeling, e.g. \cite{abc,nuxmv,ivy,sally,pono}. Although users of these systems can often overcome the tool's modeling focus by manually including support for different computation paradigms, for example, Dafny can be used to model concurrent systems \cite{dafny-concurrent}, this is not always straightforward, and limited support for different paradigms can manifest as limitations in downstream applications. For example, the Serval~\cite{serval} framework, based on Rosette, cannot reason about concurrent code. \uclid{}, to the best of our knowledge, is the only verification tool natively supporting modeling with external oracles.

\uclid{} supports different kinds of specifications and verification procedures (see Section~\ref{sec:verification_modes}). Most existing tools\cite{dafny,abc,nuxmv} do not support multi-modal verification at all. Tools that do offer multi-modal verification do not offer the same range of options as \uclid{}. For example, \cite{pono} does not support linear temporal logic, and \cite{ivy,murphi} does not support hyperproperty verification.


Finally, \uclid{} supports a generic integration with program synthesis (see Section~\ref{sec:synthesis}), and so related work includes a number of synthesis engines. 
The SKETCH system~\cite{solar2009} synthesizes expressions to fill holes in programs, and has subsequently been applied to program repair~\cite{DBLP:conf/sigsoft/LeCLGV17,DBLP:conf/icse/HuaZWK18}. \uclid{} is more flexible than this work, and allows users to declare unknown functions even in the verification annotations, as well as supporting multiple verification algorithms and types of properties.
Rosette~\cite{rosette} provides support for synthesis and verification, but, unlike \uclid{}, the synthesis is limited
to bounded specifications of sequential programs and external synthesis engines are not supported.
Synthesis algorithms have been used to assist in verification tasks, such as safety and termination of loops~\cite{DBLP:conf/lpar/DavidKL15}, and generating invariants~\cite{DBLP:conf/tacas/FedyukovichB18,DBLP:conf/vmcai/ZhangYFGM20}, but none
of this work to-date integrates program synthesis fully into an existing verification tool. 
Before the new synthesis integration, \uclid{} supported synthesis of inductive invariants. The key insight of this work is to generalize the synthesis support, and to unify all synthesis tasks by re-using the verification back-end.



\section{Software Project}
\label{sec:software}
The source code for \uclid is made publicly available under a BSD-license~\footnote{\url{https://github.com/uclid-org/uclid}}. \uclid is maintained by the \uclid team~\footnote{\url{https://github.com/uclid-org/uclid/blob/master/CONTRIBUTORS.md}}, and we welcome patches from the community. 
Additional front-ends are available for \uclid, including translators from Firrtl~\cite{firrtl1}~\footnote{\url{https://github.com/uclid-org/chiselucl}}, and RISC-V binaries~\footnote{\url{https://github.com/uclid-org/riscverifier}} to \uclid models. An artifact incuding the code for the case studies in this paper is available~\cite{artifact}.

\paragraph{Acknowledgments:} The UCLID5 project is grateful for the significant contributions by Pramod Subramanyan, one of the original creators of the tool. 
This work was supported in part by NSF grant 1837132, the DARPA grant FA8750-20-C-0156 (LOGiCS), by the Qualcomm Innovation Fellowship, and by Amazon and Intel.
\clearpage
\bibliographystyle{splncs04}
\bibliography{paper}

\appendix

\section*{Appendix}
\section{Further examples}
\subsection{Case Study: Checking reachability properties in reactive embedded systems}
\label{sec:LF}
\begin{figure}
    \centering
    \includegraphics[width=\textwidth/2]{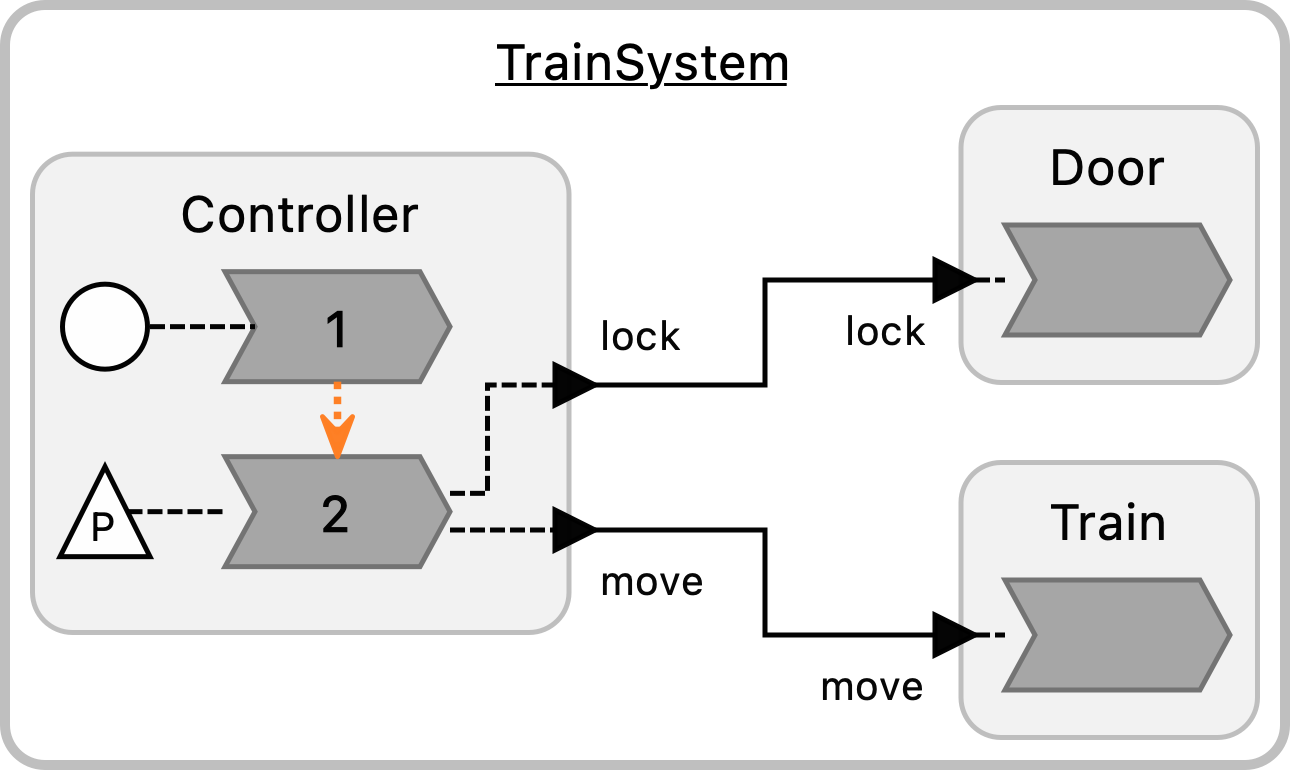}
    \caption{A simple train system.}
    \label{fig:lf_traindoor}
\end{figure}

We further demonstrate the modeling flexibility of \uclid via a case study of checking reachability properties in reactive embedded systems written in a coordination language called Lingua Franca (LF), which allows users to compose reactive components called \textit{reactors}~\cite{Lohstroh:2019:CyPhy,LohstrohEtAl:21:Towards}. LF adopts discrete event semantics in which events are processed in timestamp order. Blocks of application code, named \textit{reactions} (denoted in chevrons), can be activated by \textit{triggers}, which include startup (circle), physical action (triangle labeled by ``P''), and ports (dark solid triangle). Figure~\ref{fig:lf_traindoor} shows the diagram of a train door system with three reactors (\texttt{Controller}, \texttt{Door}, and \texttt{Train}). The driver pressing a button provides the physical action, which triggers reaction 2 in \texttt{Controller}. The reaction then outputs signals to close the door (by triggering the reaction in \texttt{Door}) and to move the train (by triggering the reaction in \texttt{Train}). Using a combination of axiomatic and operational modeling enabled by \uclid, we can  check whether the system permits an unsafe behavior where the train moves before the door closes.

\begin{figure}[h]
\begin{lstlisting}[language=uclid,style=uclidstyle]
// All timestamps and microsteps are nonnegative.
axiom(pi1(g(state)) >= 0 && pi2(g(state)) >= 0);
// Each state can either be NULL or a specific id.
axiom(id(ctrl1State) == NULL || id(ctrl1State) == controller_1);
axiom(id(ctrl2State) == NULL || id(ctrl2State) == controller_2);
axiom(id(trainState) == NULL || id(trainState) == train_1);
axiom(id(doorState)  == NULL || id(doorState)  == door_1);
next {
    havoc state; // Update state
    case
        (id(state) == door_1) : { doorState' = state; doorCloses' = true; }
        (id(state) == train_1) : { trainState' = state; trainMoves' = true; }
        (id(state) == controller_1) : { ctrl1State' = state; }
        (id(state) == controller_2) : { ctrl2State' = state; }
    esac
    // Time tags do not decrement.
    assume(tag_same(g(state'), g(state)) || tag_later(g(state'), g(state)));
    // An event triggers only once in an instant.
    assume(g(state) == g(state') ==> id(state) != id(state'));
    // Reaction with higher priority triggers first.
    assume((g(state) == g(state') && same_reactor(id(state), id(state'))) ==> priority(id(state)) < priority(id(state')));
    // Connection delay
    assume((id(ctrl1State) != NULL && id(ctrl2State') != NULL)
        ==> tag_diff(g(ctrl2State'), g(ctrl1State)) == zero());
    assume((id(ctrl2State) != NULL && id(trainState') != NULL)
        ==> tag_diff(g(trainState'), g(ctrl2State)) == zero());
    assume((id(ctrl2State) != NULL && id(doorState') != NULL)
        ==> tag_diff(g(doorState'), g(ctrl2State)) == zero());
    // Trigger mechanism
    assume(id(ctrl1State) == NULL ==> id(ctrl2State') == NULL);
    assume(id(ctrl2State) == NULL ==> id(trainState') == NULL);
    assume(id(ctrl2State) == NULL ==> id(doorState') == NULL);
}
property p: !(trainMoves && !doorCloses);
control {
    v = induction;
    check;
    print_results;
    v.print_cex;
}
\end{lstlisting}
  \caption{The \uclid model for the train system in Figure~\ref{fig:lf_traindoor}}
  \label{fig:lfUclid}
\end{figure}
The \uclid snippet in Figure~\ref{fig:lfUclid} illustrates this hybrid modeling approach. The axiomatic segment sits above the \texttt{next} block (line 1-7), specifying the semantics of reactors that should hold throughout the execution. Inside the \texttt{next} block, the \texttt{havoc} statement (line 9) sets the \texttt{state} variable in the next transition to a nondeterministic value. The \texttt{case} block (line 10-23) stores the new states in the appropriate state variables and sets boolean flags \texttt{doorCloses} and \texttt{trainMoves}. The \texttt{assume} statements (line 24-39) constrain the next nondeterministic value of \texttt{state} to one that complies with the language semantics. The operational modeling using the \texttt{next} block simplifies the specification of constraints including non-decreasing time tags (line 25), unique event per tag (line 27), reaction priority (line 29), connection delay (line 31-36), and trigger mechanism (line 38-40). The reachability property, ``the train does not move when the door is open,'' can then be checked using proof by induction (line 44).

\subsection{Case Study: Fully axiomatic modeling}
\label{sec:axiom}

\begin{figure}
\centering
\begin{subfigure}{.5\textwidth}
    \centering
    \begin{tabular}{|c|c|}
      \hline
      \textbf{Core 0}       & \textbf{Core 1} \\\hline
      (i1) [x] $\leftarrow$ 1 & (i3) r1 $\leftarrow$ [y]  \\
      (i2) [y] $\leftarrow$ 1 & (i4) r2 $\leftarrow$ [x] \\\hline
      \multicolumn{2}{|c|}{SC forbids: r1=1, r2=0} \\\hline
    \end{tabular}
    \vspace{-5pt}
    \caption{Code for litmus test \texttt{mp}.}
    \label{fig:mp_code}
\end{subfigure}%
\begin{subfigure}{.5\textwidth}
    \centering
    \includegraphics[width=0.6\textwidth]{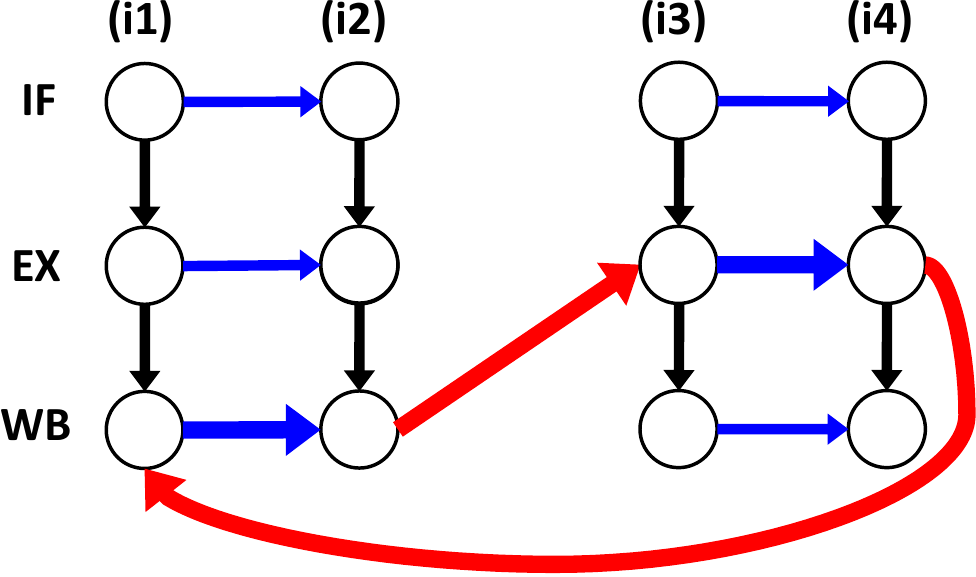}
    \caption{$\mu$hb graph for the execution of \texttt{mp}}
    \label{fig:uhbgraph}
\end{subfigure}
\caption{Left: Code for litmus test \texttt{mp}. The outcome \texttt{r1=1,r2=0} is forbidden under sequential consistency (SC)~\cite{lamport:how} (i.e., interleaving semantics). Right: An example $\mu$hb graph for the execution of the \texttt{mp} litmus test where \texttt{r1=1,r2=0} on a microarchitecture with three-stage in-order pipelines. The graph is cyclic (as highlighted by the bolded edges), implying that this execution is unobservable on the microarchitecture.}
\label{fig:test}
\vspace{-10pt}
\end{figure}


In the rest of this section, we provide an example of how to encode fully axiomatic models in \uclid, specifically the $\mu$spec specifications of COATCheck~\cite{lustig:coatcheck}.

Program executions on microarchitectures (component-level models of hardware) can be represented as microarchitectural happens-before ($\mu$hb) graphs~\cite{lustig:pipecheck}.
Figure~\ref{fig:uhbgraph} depicts an example $\mu$hb graph for the execution of the \texttt{mp} litmus test~\footnote{Litmus tests are small 4-8 instruction programs used in the verification of memory consistency~\cite{adve:tutorial}. $\mu$hb graphs and $\mu$spec specifications are typically used for memory consistency verification, but they can also be used for hardware security verification~\cite{trippel:checkmate}.} on a pedagogical microarchitecture with three-stage in-order pipelines of Fetch (\texttt{IF}), Execute (\texttt{EX}), and Writeback (\texttt{WB}) stages.
Nodes in these graphs represent sub-events in instruction execution. For instance, the first node in the second row represents the event when instruction \texttt{i1} from \texttt{mp} performs its Execute stage.
Meanwhile, edges in $\mu$hb graphs represent happens-before relationships. For instance, the blue edge between the first two nodes in the second row enforces that instruction \texttt{i1}'s Execute stage must occur before the Execute stage of instruction \texttt{i2}, reflecting the in-order nature of this processor's pipelines.

The presence or absence of nodes and edges in $\mu$hb graphs for a given microarchitecture are enforced by \emph{axioms} in the domain-specific language $\mu$spec~\cite{lustig:coatcheck}. $\mu$spec supports propositional logic over its built-in predicates. It also supports quantifiers over instructions and enforces that the set of edges in $\mu$hb graphs is closed under transitivity.

We embedded $\mu$spec into \uclid to showcase \uclid's capability for axiomatic modeling, as well as to enable $\mu$spec models to benefit from \uclid's built-in capabilities for modularity and synthesis (Section~\ref{sec:synthesis}). Figure~\ref{fig:uhbUclid} shows part of this embedding as well as an example $\mu$spec axiom written using the embedding.

\begin{figure}[t]
\begin{lstlisting}[language=uclid,style=uclidstyle]
type uhbNode_t = record { nExists : boolean, nTime : integer };

type microop_t = record {
    globalID : integer, coreID : integer,
    <...>,
    Fetch : uhbNode_t, Execute : uhbNode_t,
    Writeback : uhbNode_t
};

define EdgeExists (src, dest : uhbNode_t) : boolean =
 (src.nExists == true && dest.nExists == true && src.nTime < dest.nTime);
define NodeExists (n : uhbNode_t) : boolean = (n.nExists == true);
define ProgramOrder (i, j : microop_t) : boolean = 
 (i.globalID < j.globalID && i.coreID == j.coreID);

var i1, i2, i3, i4 : microop_t;
group testInstrs : microop_t = {i1, i2, i3, i4};

axiom ex_in_order :
 finite_forall (a : microop_t) in testInstrs ::
 finite_forall (b : microop_t) in testInstrs ::
  EdgeExists(a.Fetch, b.Fetch) ==> EdgeExists(a.Execute, b.Execute);
\end{lstlisting}
  \vspace{-10pt}
  \caption{Part of the embedding of $\mu$spec in \uclid and an example $\mu$spec axiom written in this embedding, illustrating \uclid's capability for axiomatic modeling.}
  \vspace{-15pt}
  \label{fig:uhbUclid}
\end{figure}

We represent a $\mu$hb node in \uclid using two variables: a Boolean variable to represent whether or not the node exists and an integer recording the execution timestamp at which it occurred. An instruction (\texttt{microop\_t}) consists of the nodes representing its sub-events as well as metadata such as its global ID (a unique identifier), core ID, address, and data value (some fields are not shown for brevity).

The existence of an edge (\texttt{EdgeExists}) between two nodes can be modeled as enforcing that both the source and destination nodes exist, and constraining the source node's timestamp to be less than that of the destination node.
Node existence (\texttt{NodeExists}) merely checks the value of the \texttt{nExists} variable, while \texttt{ProgramOrder} is determined by ascending order of \texttt{globalID} on the same core.

The axiom \texttt{ex\_in\_order} states that for every pair of instructions \texttt{a} and \texttt{b} in the group \texttt{testInstrs}, if an edge exists between their \texttt{IF} stages, then an edge must also exist between their \texttt{EX} stages. Thus, this axiom enforces the existence of the blue edges between the \texttt{EX} stages of \texttt{i1} and \texttt{i2} and between those of \texttt{i3} and \texttt{i4} in Figure~\ref{fig:uhbgraph}.

Since edges in $\mu$hb graphs represent happens-before relationships, a cyclic $\mu$hb graph implies that an event must happen before itself. Thus, a cyclic $\mu$hb graph represents an execution that is unobservable (i.e., impossible) on the microarchitecture being modeled. Likewise, an acyclic $\mu$hb graph represents an execution that is observable on the microarchitecture. A given litmus test outcome can be verified on a microarchitecture by grounding the axioms over the instructions and outcome of that litmus test and asking a SMT solver to search for an acyclic $\mu$hb graph satisfying the axioms. If the solver returns a satisfying assignment, the test outcome is observable on the microarchitecture. If the solver returns UNSAT, the test outcome is guaranteed to be unobservable on the microarchitecture.

While litmus test verification of $\mu$spec specifications using SMT-based approaches has been conducted by prior work~\cite{lustig:coatcheck}, encoding such modeling in \uclid has the benefit of harnessing \uclid's built-in capabilities for modularity and synthesis. In fact, we are currently using \uclid's synthesis capability in ongoing work to synthesise $\mu$spec axioms that match a set of examples.

\subsection{Example: Control Synthesis with External Oracles}
\label{sec:control}
Consider the task of synthesising a controller for a Linear Time Invariant similar to Abate et al.~\cite{control-cegis}. We use a state-space representation, which is discretized in time.
$
\label{eq:ode1}
\dot{x}_{t+1} = \mat{A}\vec{x}_t+ \mat{B} \vec{u}_t, 
$
where $\vec{x} \in \mathbb{R}^n$,  
$\vec{u} \in \mathbb{R}^p$ is the input to the system, calculated as $\mat{K}\vec{x}$ where $\mat{K}$ is the controller to be synthesized, 
$\mat{A} \in \mathbb{R}^{n \times n}$ is the system matrix, 
$\mat{B} \in \mathbb{R}^{n \times p}$ is the input matrix,
and subscript $t$ indicates the discrete time step. 

We aim to find a stabilizing controller $\mat{K}$, such that absolute values of the (potentially complex) eigenvalues of the closed-loop matrix 
$\mat{A} - \mat{B}\mat{K}$ 
are less than one, checked by the oracle function \codelike{isStable}. We further require that the controller guarantees the states remain within a safe region of the state space up to a given number of time steps, using the bounded model checking verification command in UCLID5, as shown in Figure~\ref{ex:control}.

\begin{figure}[h]
\begin{lstlisting}[language=uclid,style=uclidstyle]
module main {
  var x0, x1: float;
  group stateVars : float = {x0, x1};
  const a00, a01, a10, a11 : float;
  const b0, b1 : float;
  const AminusBK00, AminusBK01, AminusBK10, AminusBK11 : float;
  oracle function [isstable] isStable(s00:float, s01:float, s10:float, s11:float) : boolean;
  synthesis function k0 (): float;
  synthesis function k1 (): float;

  // LTI system spec
  axiom A: (a00==0.901224922471 && a01==0.000000013429 && a10==0.000000007451 && a11==0.000000000000);
  axiom B: (b0==128.000000000000 && b1==0.000000000000);
  axiom ax1: AminusBK00 == a00 - b0*k0() && !isNaN(AminusBK00);
  axiom ax2: AminusBK01 == a01 - b0*k1() && !isNaN(AminusBK01);
  axiom ax3: AminusBK10 == a10 - b1*k0() && !isNaN(AminusBK10);
  axiom ax4: AminusBK11 == a11 - b1*k1() && !isNaN(AminusBK11);

  init { // bound initial states
    assume (finite_forall (state: float) in stateVars :: state<0.1 && state > -0.1);
  }
  next { // step the system
    x0' = AminusBK00*x0 + AminusBK01*x1;
    x1' = AminusBK10*x0 + AminusBK11*x1;
  }
  // the safety condition
  invariant stability: isStable(AminusBK00, AminusBK01, AminusBK10, AminusBK11);
  invariant safety: finite_forall (state: float) in stateVars :: state < 1.0 && state > -1.0;
  invariant isNotNaN: finite_forall (state: float) in stateVars :: state < 1.0 && state > -1.0;
 
  control {
    unroll(10); // fix safety bound
    check;
  }
}
\end{lstlisting}
  \caption{\uclid control synthesis example. The next block assigns to the state variables according to the standard definition of Linear Time Invariant systems. Note this model uses finite quantifiers, as described in Section~\ref{sec:operAx}.}
  \label{ex:control}
\end{figure}


\end{document}